# Chromospheric Anemone Jets as Evidence of Ubiquitous Reconnection

Kazunari Shibata,[1*] Tahei Nakamura,[1] Takuma Matsumoto,[1] Kenichi Otsuji,[1] Takenori J. Okamoto,[1,2] Naoto Nishizuka,[1] Tomoko Kawate,[1] Hiroko Watanabe, [1] Shin'ichi Nagata,[1] Satoru UeNo,[1] Reizaburo Kitai,[1] Satoshi Nozawa,[3] Saku Tsuneta,[2] Yoshinori Suematsu,[2] Kiyoshi Ichimoto,[2] Toshifumi Shimizu,[4] Yukio Katsukawa,[2] Theodore D. Tarbell,[5] Thomas E. Berger, [5] Bruce W. Lites,[6] Richard A. Shine,[5] Alan M. Title[5]

[1]Kwasan and Hida Observatories, Kyoto University, Yamashina, Kyoto, 607-8471, Japan.

[2]National Astronomical Observatory, Mitaka, Tokyo, 181-8588, Japan.

[3]Department of Science, Ibaraki University, Mito, Ibaraki, 310-8512, Japan.

[4]Institute of Space and Astronomical Science/ Japan Aerospace Exploration Agency, Sagamihara, Kanagawa, 229-8510, Japan.

[5]Lockheed Martin Solar and Astrophysics Laboratory, B/252, 3251 Hanover Street, Palo Alto, CA 94304, USA.

[6]High Altitude Observatory, National Center for Atmospheric Research, Post Office Box 3000, Boulder CO 80307-3000, USA
*To whom correspondence should be addressed. E-mail: shibata@kwasan.kyoto-u.ac.jp

**The heating of the solar chromosphere and corona is a long-standing puzzle in solar physics. Hinode observations show the ubiquitous presence of chromospheric anemone jets outside of sunspots in active regions. They are typically 3 τo 7 arcsec = 2000 to 5000 kilometers long and 0.2 to 0.4 arcsec = 150 to 300 kilometers wide, and their velocity is 10 to 20 kilometers per second. These small jets have an inverted Y-shape, similar to the shape of x-ray anemone jets in the corona. These features imply that magnetic reconnection similar to that in the corona is occurring at a much smaller spatial scale throughout the chromosphere and suggest that the heating of the solar chromosphere and corona may be related to small scale ubiquitous reconnection.**

Recent solar space missions such as Yohkoh, SOHO, TRACE, and RHESSI have revealed that the outer atmosphere of the Sun, such as the solar corona and transition region, is much more dynamic than had been thought and contains numerous jets and microflares. The mechanism providing such a highly dynamic corona and transition region is a mystery and may be related to the long-standing puzzle of how the corona is heated to a million K (*1*). It has long been observed with H $\alpha$ observations from the ground that cool jets, called surges, often occur in the chromosphere near sunspots in association with small flares (*2*). Here we report the discovery of tiny chromospheric anemone jets with the Solar Optical Telescope (SOT) aboard Hinode (*3*). These jets are much smaller and occur much more frequently than surges and are considered to be indirect evidence of small scale ubiquitous reconnection in the solar atmosphere as conjectured by Parker (*4*).

Figure 1 shows the Ca II H broadband filter snapshot image of the active region on 17 December 2006 near the west limb, taken with Hinode/SOT (see also movie S1) The dark elongated ellipse seen near the limb is a sunspot. Numerous tiny jets ejected from bright points (arrows) can be seen. There are also numerous jets or jetlike structures throughout this image, some of which are spicules, fibrils, and microjets above sunspot penumbra (*5*). We concentrate on the isolated jets from bright points, which we call Ca jets. Figure 2 shows the time evolution of four typical Ca jets. The footpoints of these jets are not a simple bright point, but show a cusp or inverted Y-shape. This is the characteristic shape of x-ray jets, known as anemone jets (*6*-*8*).

On the basis of Yohkoh observations (*6*), the anemone shape is formed as a result of magnetic reconnection between an emerging magnetic bipole and a preexisting uniform vertical field. Namely, once reconnection occurs between the emerging bipole and vertical field, the field lines with a polarity opposite to that of the ambient field become connected to the ambient polarity regions, forming a fan shape like that of a sea anemone (*6*). Hence, this kind of structure has been called an anemone. X-ray jets discovered by Yohkoh, tend to have an anemone shape at their footpoint (*9*), providing indirect evidence of magnetic reconnection as a mechanism for jet formation. Most of the footpoints of x-ray jets correspond to mixed polarity regions (*10*), supporting a magnetic reconnection mechanism. Two-dimensional magnetohydrodynamic (MHD) numerical simulations of reconnection between emerging flux and uniform coronal field also reproduced well the observed characteristics of x-ray jets (*7*, *8*). In this case, the reconnection occurs in the low corona or in the upper chromosphere, producing high-speed jets. Because the characteristic velocity of a reconnection jet is the local Alfvèn speed, the velocity of the high-speed jet becomes comparable to the coronal Alfvèn speed, or 100 to 1000 km $s^{-1}$ or more, which is comparable to the actual observed velocity of x-ray jets. The plasma in the corona is heated to temperatures ranging from a few million K to about 10 million K. This hot plasma can be observed as microflares and soft x-ray jets (Fig. 3A). At the same time, cool jets are accelerated if the reconnection occurs in the upper chromosphere where cool ($10^4$ K) plasma is situated near the reconnection point. These cool jets may correspond to H $\alpha$ surges, and indeed, observations confirm the coexistence of x-ray jets and H $\alpha$ surges (*11*, *12*, *13*).

Because magnetic reconnection is a universal physical process that can occur on any spatial or temporal scale (*14*), it should occur in the chromosphere and the photosphere (15), as well as the corona. The extreme-ultraviolet jets (*16*) and H $\alpha$ surges (2, 17) may be produced in this way in configurations like those shown in Fig. 3. If a tiny emerging bipole appears, or if a tiny bipole collides with an ambient field of opposite polarity, we can expect magnetic reconnection to occur at a relatively small spatial scale in a manner similar to that of an anemone jet observed in Yohkoh soft x-rays (*18*) (Fig. 3C). Hence the observed morphology of the Ca jets, such as the inverted Y-shape or anemone jet, suggests that they are revealing magnetic reconnection. [Compare the expected magnetic-field configuration and jet based on a reconnection model (Fig. 3D) with that of actual observation (Fig. 3E).] Because the plasma density is high in the low chromosphere and photosphere, the Alfvèn speed is low, on the order of 10 km $s^{-1}$, which is comparable to the speed of sound in those layers. The plasma temperature increases by only a factor of $\leqq 2$ because the ratio of gas pressure to magnetic pressure ($\beta$) is near unity and the radiative cooling time is short in the low chromosphere and photosphere. This is consistent with the observations of Ca jets.

Figure 4 shows that there are multiple jet events in the same location and that the apparent (projected) velocity of Ca jets is typically 10 to 20 km $s^{-1}$, which is the same range of values predicted for cool jets in the low chromosphere and photosphere (*18*). That is, the Alfvèn velocity in the low chromosphere is estimated to be $V_A \approx 10$ km $s^{-1}$ $(B/100\ G)\ (n/10^{15}\ cm^{-3})^{-1/2}$, if we assume that the density in the low chromosphere is $10^{15}$ $cm^{-3}$ and the magnetic field strength is 100 G. The relation between the Ca jets and the surrounding magnetic field distribution also shows that some of Ca jets occur in mixed polarity regions, or near the boundary between opposite polarities (fig. S1), supporting the magnetic reconnection mechanism.

The total energy involved in a Ca jet and the footpoint is not easy to determine because the density of Ca jets is difficult to estimate. If we assume that the density at the footpoint of Ca jets is $10^{15}$ $cm^{-3}$ and its temperature is 5000 K, the total thermal energy content is $E_{\rm thermal} \approx (3/2)nkTd^3 \approx 3 \times 10^{25}$ erg $(n/10^{15}\ cm^{-3})\ (T/5000K)\ (d/300\ km)^3$, where $d$ is the typical size of the footpoint (bright point), about 0.4 arcsec ≈ 300 km from our observations. The total magnetic energy stored in the same volume is $E_{mag} \approx (B^2/8\pi)d^3 \approx$ $10^{25}$ erg $(B/100G)^2\ (d/300\ km)^3$. Hence the total stored energy at the footpoint is less than a microflare value ($10^{26}$ erg), and smaller fluctuations occurring at the footpoint would correspond to a smaller release of energy, $10^{23}$ to $10^{24}$ erg, lying in the nanoflare regime.

The velocity of the Ca jets is typically 10 to 20 km $s^{-1}$. But if the jet undergoes ballistic motion, i.e., if the jet is decelerated by gravity, the maximum height of the jet would be $H_{jet,\rm max} \approx V_{jet}^2/(2g) \approx 200$ km $(V_{jet}/10\ km\ s^{-1})^2$, which is much smaller than the actual height of the jet, 2000 to 5000 km. Hence the jet cannot reach the observed height if it undergoes ballistic motion. So how can we solve this puzzle ? There are two possibilities. One is that the jet is continuously accelerated beyond the height of 200 km. If the jet is situated along the current sheet, this would be possible. However, it is unlikely that jets with a length longer than ~ 1 arcsec (~ 700 km), such as those shown in Fig. 2, are along a current sheet longer than 1 arcsec because the bipoles at the footpoint of the jets are usually smaller than 1 arcsec (see also an example of a magnetogram in fig.

S1) and because the mixed polarities at the footpoint of the jet are often formed as a result of an emerging bipole. The other possibility is that the main component of the observed Ca jet is not the plasma ejected from the footpoint but the plasma ejected from the upper chromosphere as a result of the propagation of slow mode magnetoacoustic shocks or nonlinear Alfvèn waves (or fast-mode shocks) (Fig. 3C). Shibata *et al*. (*19*) showed that when the explosion occurs below the middle chromosphere, even if it is small so that the plasma ejected from the explosion cannot reach higher altitude, a slow mode shock is formed ahead of the ejecta and grows substantially. The shock accelerates the plasma in the upper chromosphere, which can reach a height of 3000 to 10000 km. Spicules and small surges ejected from Ellerman bombs may be generated in this way (*20*). The nonlinear Alfvèn wave can play a similar role to accelerate the jet from the upper chromosphere (*21–23*).

Parker (*4*) proposed that nanoflares (reconnection) heat the corona in quiet and active regions, and that nanoflares generate Alfvèn waves, which eventually accelerate high-speed solar winds in coronal holes (*24*, *25*). MHD simulations (*26*, *27*) showed that about 10 % of energy released in the reconnection is carried away by Alfvèn waves and a comparable amount by slow mode magnetoacoustic shocks in a typical geometry. Our finding of numerous Ca jets in the low chromosphere with indirect evidence of reconnection is consistent with these ideas.

Are these Ca jets enough to heat the corona in these regions ? The total number of Ca jets in the field of view of Fig. 1 during a 1-hour period (20:00 to 21:00 UT on 17 December 2007) was 59. The total area of this field of view is roughly 50,000 $km^2$, so that the Ca jets occur at a rate of $7 \times 10^{-22}$ $s^{-1}$ $cm^{-2}$. Combining this value with the typical energy of a Ca jet of $\sim 10^{25}$ erg, the average energy flux carried by these Ca jets is $7 \times 10^3$ erg $cm^{-2}$ $s^{-1}$. This is much smaller than the necessary energy flux to heat the active region corona, $\sim 10^6$ erg $cm^{-2}$ $s^{-1}$. However, by definition, a Ca jet has a clear bright footpoint with an anemone shape, and a movie of the Ca (movie S1, corresponding to Fig. 1) shows that almost all parts of the active-region chromosphere are covered by numerous tiny jets or jetlike features, whose footpoints are not well resolved even with Hinode. These jets seem to be continuously related to spicules, fibrils, and micro-jets above the penumbrae of sunspots (*5*), and it is likely that Ca jets are the largest events in a spectrum that has many more, smaller events. Hence these and other Hinode data (*5*, *28*) suggest that magnetic reconnection occurs throughout the solar atmosphere and may play an important role in heating it, as conjectured by Parker (*4*).

30. We thank E. N. Parker, T. Magara and S. Morita for useful comments. Hinode is a Japanese mission developed and launched by the Institute of Space and Astronomical Science/Japan Aerospace Exploration Agency, with the National Astronomical Observatory of Japan as domestic partner and NASA and Science and Technology Facilities Council (UK) as international partners. It is operated by these agencies in cooperation with the European Space Agency and Norwegian Space Centre (Norway). This work was supported in part by the Grant-in-Aid for the 21st Century Center of Excellence "Center for Diversity and Universality in Physics" from the Ministry of Education, Culture, Sports, Science and Technology of Japan, and in part by the Grant-in-Aid for Creative Scientific Research "The Basic Study of Space Weather Prediction" (head investigator: K. Shibata) from the Ministry of Education, Science, Sports, Technology, and Culture of Japan.

**Supporting Online Material**

www.science.mag.org/cgi/content/full/VOL/ISSUE/PAGE/DC1

Figure S1

Movie S1

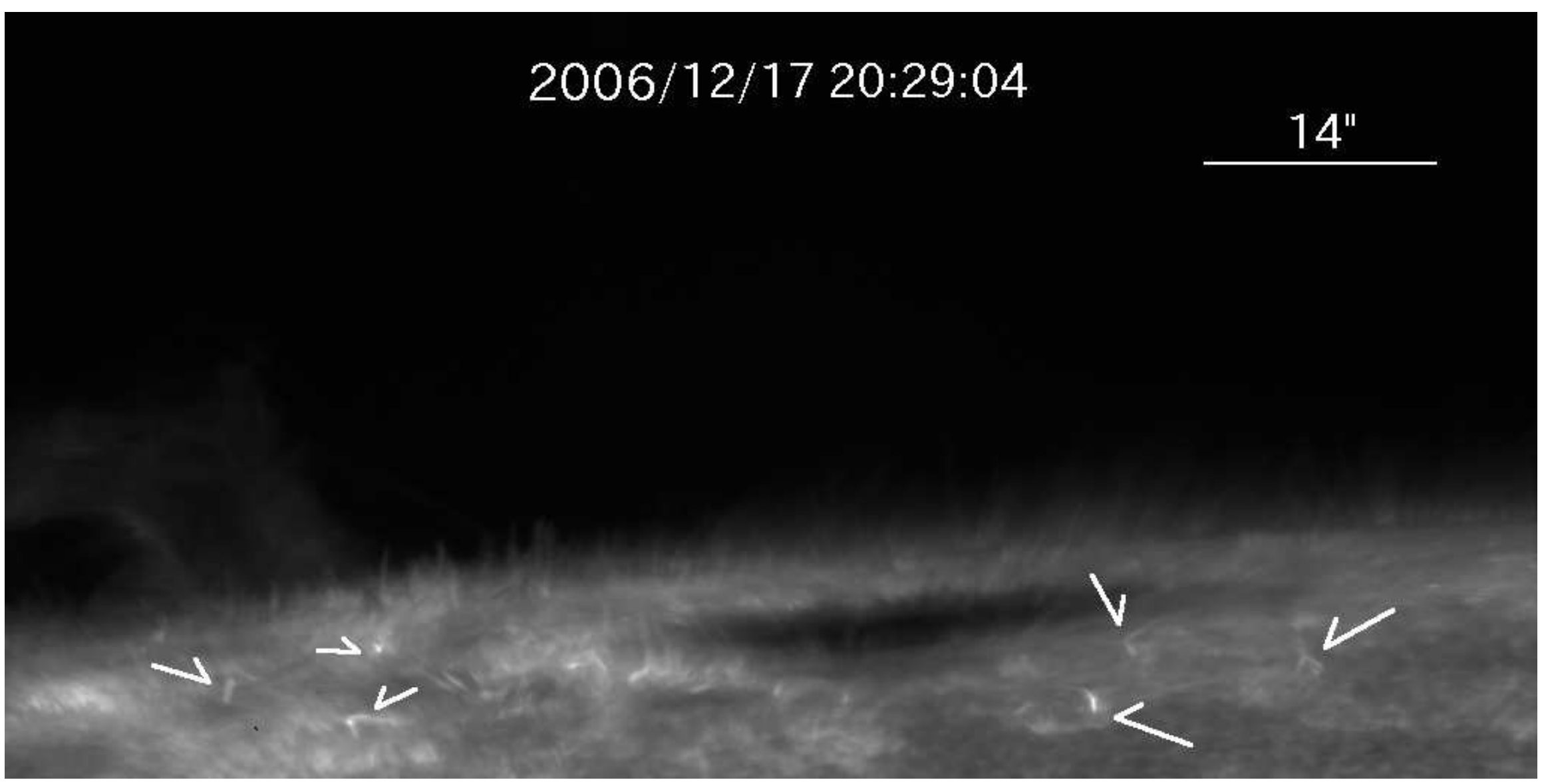

**Fig. 1.** Ca II H broad band filter image taken with the SOT aboard Hinode at 20:29:04 UT on 17 December 2006. For performance and examples of Ca II H observations with Hinode/SOT, see e.g., (29).

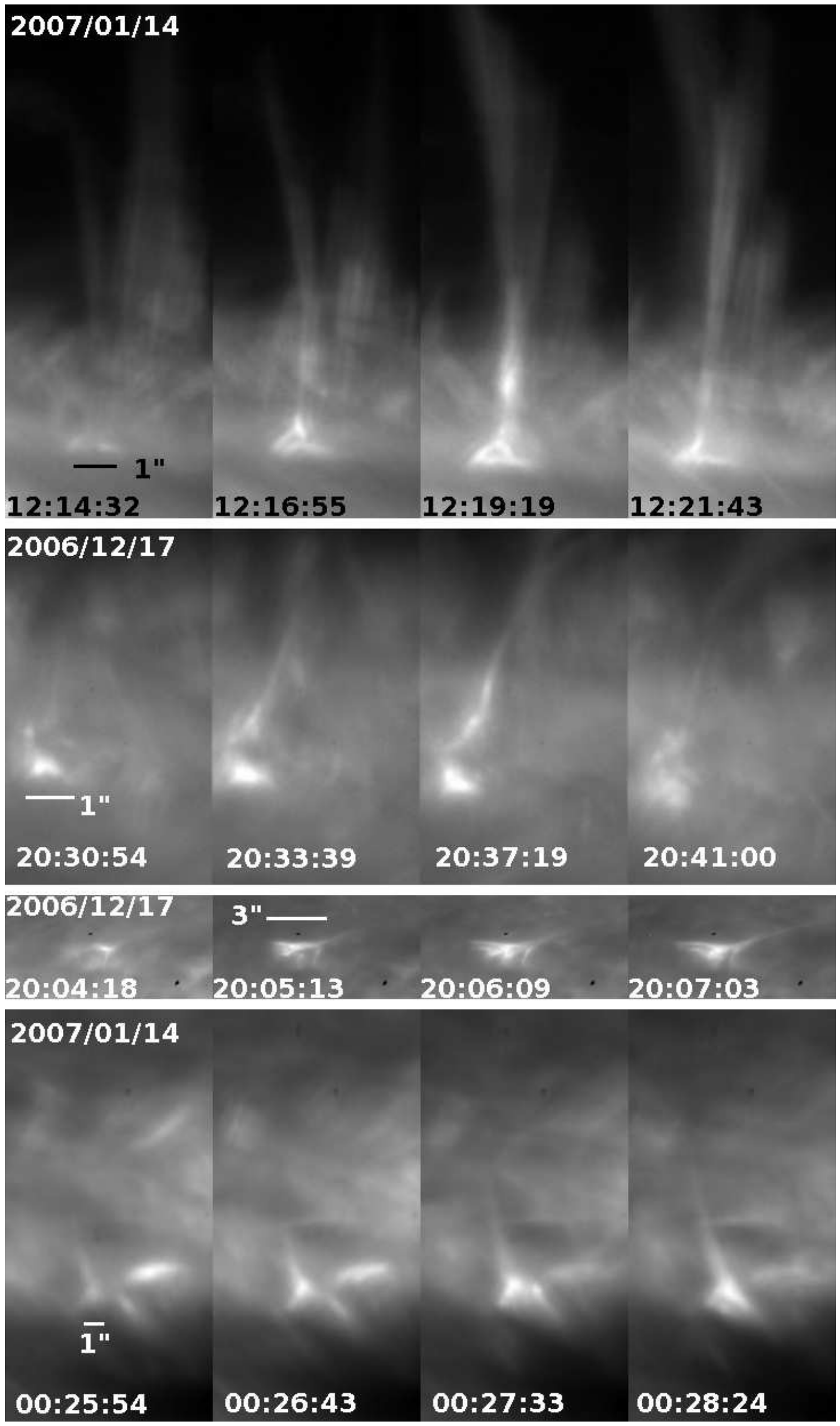

**Fig. 2.** Time evolution of typical Ca jets observed in Ca II H broad band filter of Hinode/SOT. Times are shown in UT.

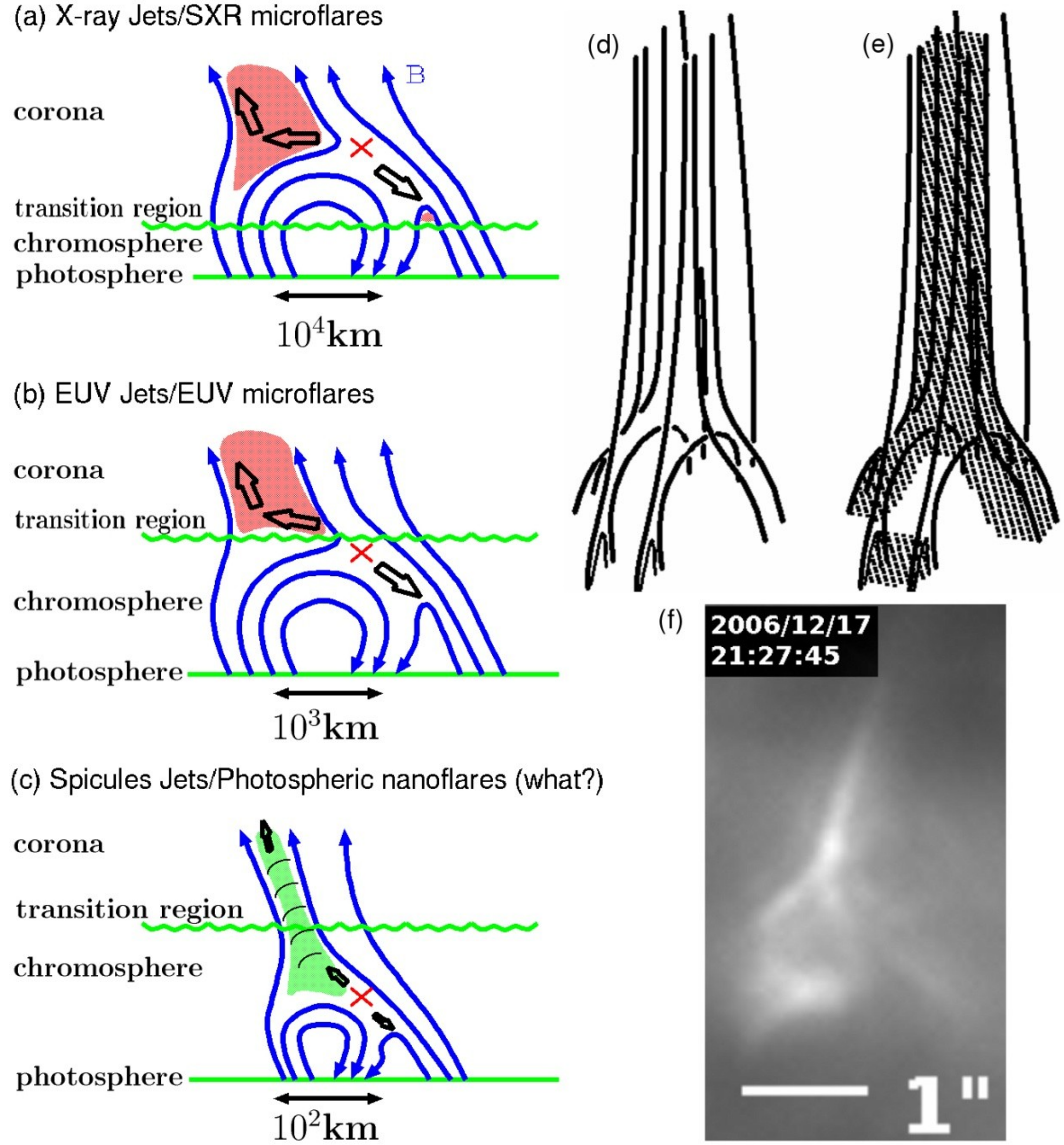

**Fig. 3.** (A to C) Schematic illustration of magnetic reconnections that occur at various altitudes [adapted from (*19*)]. (D and E) A model of Ca jets. (F) A typical example of an observed Ca jet.

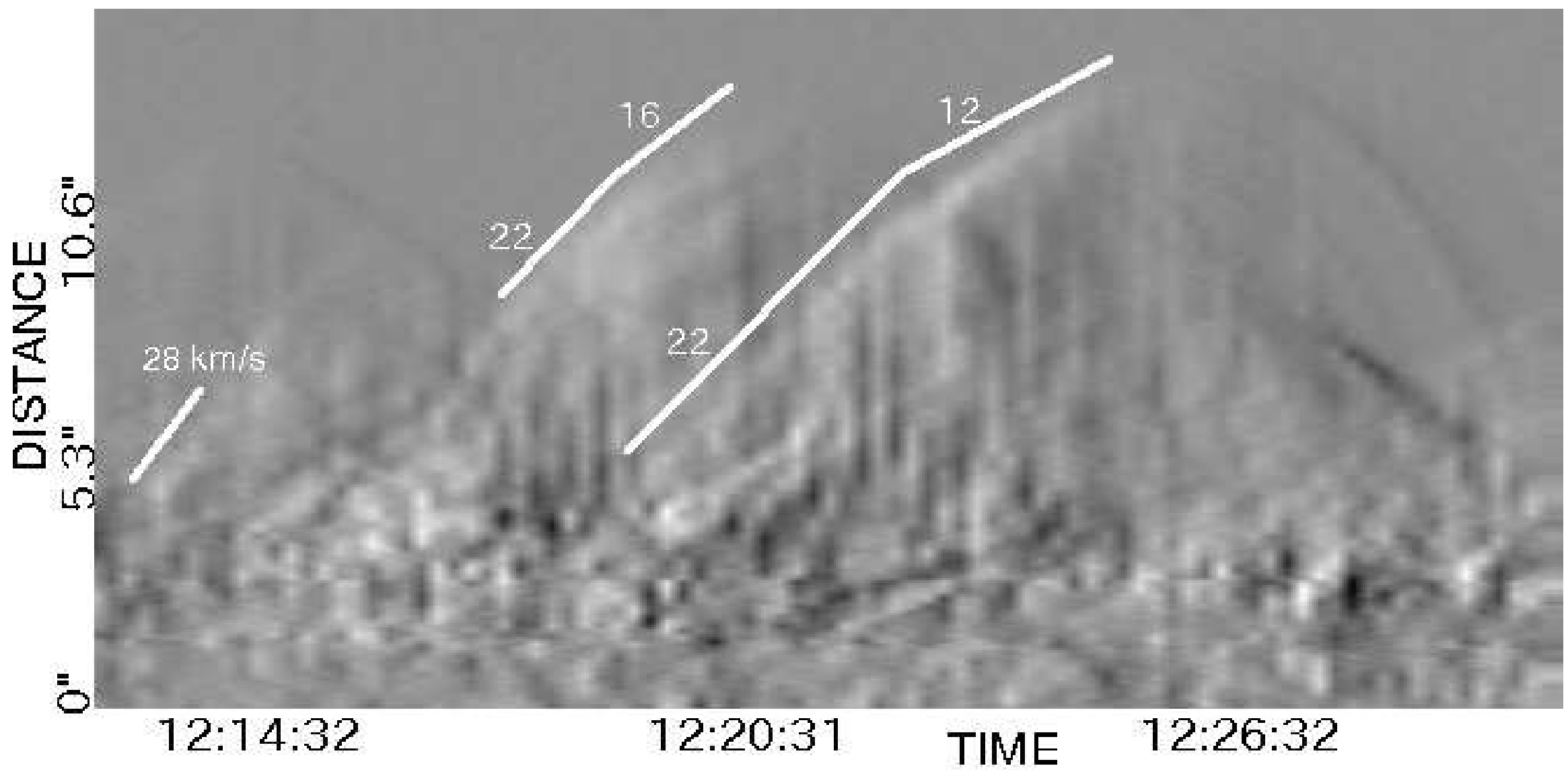

**Fig. 4.** Time-distance diagram of Ca intensity (running differences) along Ca jets, observed with the Ca II H broad-band filter of Hinode/SOT. The numbers above the curves denote velocity (in km $s^{-1}$).

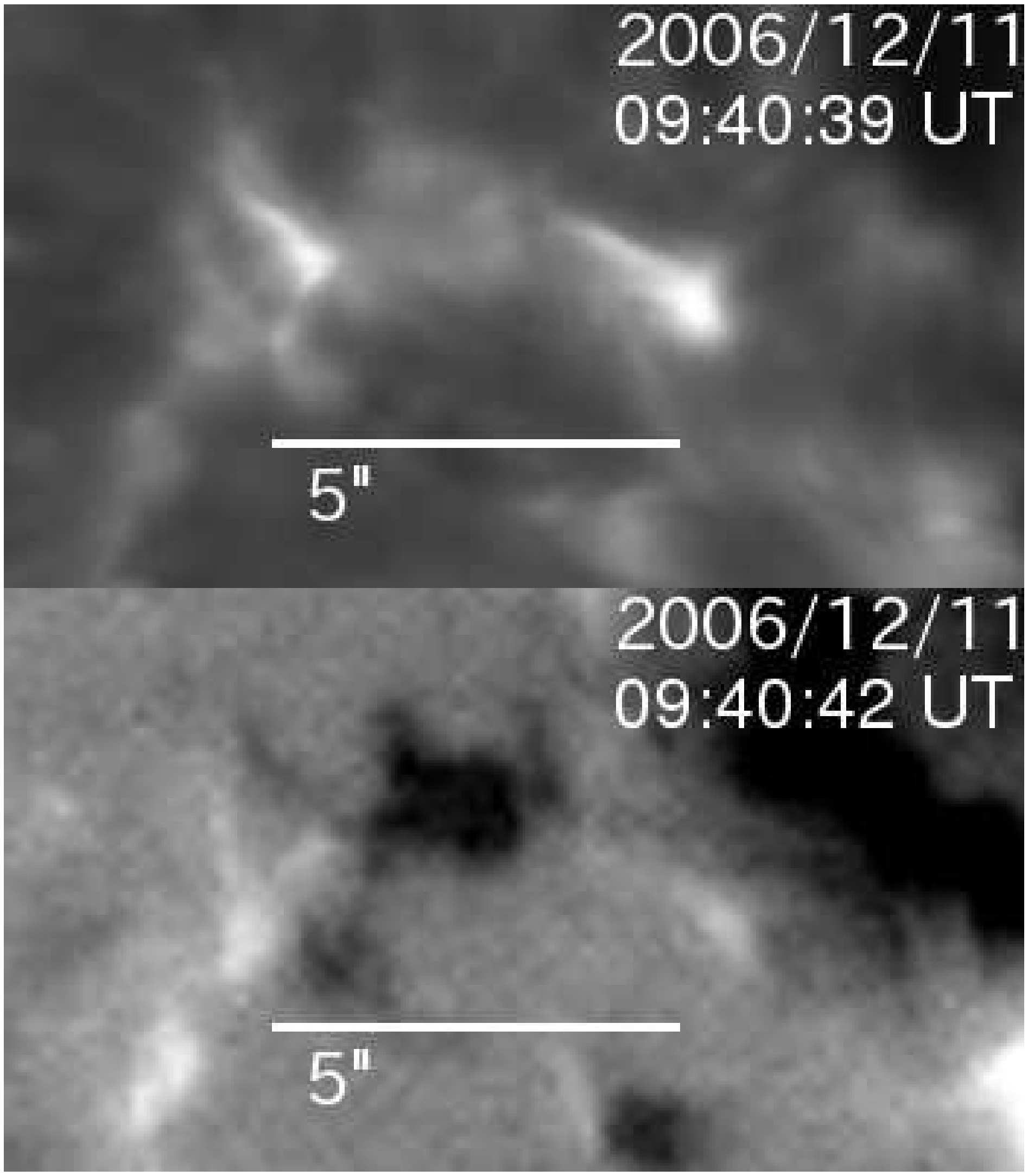

**Fig. S1.** Upper panel shows Ca II H image for two Ca jets and lower panel shows magnetogram (narrow band V filter image) at the same time and the same field of view as those of Ca jets, taken with Hinoede/SOT. These images show that there are mixed polarities at the footpoint of the Ca jets, supporting magnetic reconnection mechanism for the formation of Ca jets.